\documentclass[pra,aps,10pt, twocolumn, floatfix]{revtex4-1} %%

\usepackage {graphicx} %%
\usepackage {amssymb} %%
\usepackage {amsmath}
\begin{document}

\title{Generation of unipolar half-cycle pulse via unusual reflection of a single-cycle pulse from an optically thin metallic or dielectric layer}

\author{M.V. Arkhipov}

\affiliation{Faculty of Physics, St. Petersburg State University, Ulyanovskaya 1, Petrodvoretz, St. Petersburg 198504, Russia}

\author{R. M. Arkhipov}

\affiliation{St. Petersburg State University, Faculty of Physics, Ulyanovskaya str. 3, Petrodvorets, St. Petersburg 198504, Russia}
\affiliation{ITMO University, Kronverkskiy prospekt 49, 197101, St. Petersburg,  Russia}
\affiliation{Max Planck Institute for the Science of Light, Staudtstr. 2, 91058 Erlangen, Germany}

\author{A.V. Pakhomov}

\affiliation{Samara National Research University, Moskovskoye Shosse 34, Samara 443086, Russia}
\affiliation{Department of Theoretical Physics, Lebedev Physical Institute, Novo-Sadovaya str. 221, Samara 443011, Russia}

\author{I. Babushkin}

\affiliation{Institute of Quantum Optics, Leibniz University Hannover, Welfengarten 1, 30167 Hannover, Germany}
\affiliation{Max Born Institute, Max-Born-Strasse 2a, Berlin 10117, Germany}

\author{A. Demircan}

\affiliation{Institute of Quantum Optics, Leibniz University Hannover, Welfengarten 1, 30167 Hannover, Germany}
\affiliation{Hannover Centre for Optical Technologies, 30167 Hannover, Germany}

\author{U. Morgner}

\affiliation{Institute of Quantum Optics, Leibniz University Hannover, Welfengarten 1, 30167 Hannover, Germany}
\affiliation{Hannover Centre for Optical Technologies, 30167 Hannover, Germany}
\affiliation{ITMO University, Kronverkskiy prospekt 49, 197101, St. Petersburg,  Russia}

%\affil[*]{Corresponding author: arkhipovrostislav@gmail.com}

%\dates{Compiled \today}

%\ociscodes{(140.4050) Mode-locked lasers; (190.0190) Nonlinear optics, (020.1670) Coherent optical effects; (140.7090) Ultrafast lasers, (190.5530) Pulse propagation and temporal solitons; (270.3430) Laser theory}

%\doi{\url{http://dx.doi.org/10.1364/ao.XX.XXXXXX}}

\begin{abstract}
 We present a significantly different reflection process from an optically thin flat metallic or dielectric layer and propose a strikingly simple method to form approximately unipolar half-cycle optical pulses via reflection of a single-cycle optical pulse. Unipolar pulses in reflection arise due to specifics of effectively one-dimensional pulse propagation. Namely, we show that in considered system the field emitted by a flat medium layer is proportional to the velocity of oscillating medium charges instead of their acceleration as it is usually the case. When the single-cycle pulse interacts with linear optical medium, the oscillation velocity of medium charges can be then forced to keep constant sign throughout the pulse duration. Our results essentially differ from the direct mirror reflection and suggest a possibility of unusual transformations of the few-cycle light pulses in linear optical systems.
\end{abstract}

%\setboolean{displaycopyright}{true}
%\begin{document}

\maketitle
%\thispagestyle{fancy}
%\ifthenelse{\boolean{shortarticle}}{\abscontent}{}

Generation of optical pulses with duration of single and half optical cycle becomes now of practical and fundamental importance because number of their applications increasingly grows 
\cite{Ramasesha, Peng, Kling, Pfeifer}. In particular, it allows to probe intramolecular and even intraatomic processes, what has led in the past decades to the emergence of the new field
of attosecond spectroscopy \cite{Krausz, Baltuska, Goulielmakis}. Among the ultrashort pulses, unipolar half-cycle ones are of special interest. In contrast to oscillating
few- or single-cycle pulses, they maintain the constant sign of the electric field throughout the whole pulse duration. This specific feature opens up new opportunities for their use in 
practice which were found theoretically or demonstrated experimentally with presently realizable subpicosecond half-cycle pulses. In particular, unipolar half-cycle pulses were shown to allow
for efficient control and measurement of wavepackets dynamics through the transfer of kinetic momentum to bounded electrons in atoms and molecules \cite{Jones, Matos-Abiague, Mestayer,
Zhang2009, Pavlyukh, Jones-2, Bensky, Reinhold}, driven relativistic laser-plasma interactions and charge particles acceleration \cite{Rau, Hojo}, effective control of 
photoemitted electrons and ionization processes \cite{Zhang2006, Wetzels, Raman}, ultrafast control of coherent resonant light-matter interactions \cite{Arkhipov-gratings, Arkhipov-gratings-2}. 

Such subcycle pulses can be generated using different techniques of the waveform synthesis \cite{Hassan, Manzoni}. The subcycle synthesis techniques employed for the control of the pulse
shape require complicated and bulky setups whereas the efficiency of energy conversion of pumping to the generation field is low. Hence, looking for the simple and effective ways of unipolar
subcycle pulse synthesis and control still presents the challenging issue in modern ultrafast optics.

While the most commonly available half-cycle pulses in the terahertz range \cite{Gao} can be used for the control of  highly excited Rydberg states \cite{Jones, Wesdorp}, femtosecond and
subfemtosecond ones can be promising to extend the wavepacket control to low-leaving states. Such half-cycle pulses were directly produced using high-order harmonic generation from a 
solid target irradiated with intensive few-cycle laser pulses \cite{Wu, Ma, Li}. Theoretically, such possibility was predicted to occur when an initially bipolar ultrashort pulse propagates in a nonlinear
resonant medium \cite{Kozlov, Song, Song-2, Song-3}. Generation of unipolar dissipative solitons in a two-level resonant medium was also theoretically predicted \cite{Bullough, Kalosha, Vysotina, 
Rosanov, Vysotina-2}. Another approach, based on excitation of nonlinear oscillators by a train of few-cycle pump pulses was proposed in \cite{Arkhipov, Arkhipov-2, Arkhipov-3, Pakhomov, 
Pakhomov-PRA}. 

In this paper we describe another method which we expect to be easier for applications in practice. We consider an optically thin flat layer of the dielectric or metal  irradiated by a single-cycle pulse of the plane wavefront. We demonstrate that such flat layer can exhibit unusual reflection properties and under certain conditions an incident single-cycle pulse would be transformed into a unipolar half-cycle pulse when reflected from the layer. Importantly, no nonlinear response is needed for such transformation. This approach allows to produce nearly half-cycle pulses which durations can extend into the femtosecond range. It is worthwhile noting here the
significant progress made in recent times in production of ultrathin crystalline films, including single layer or 2D topological materials, mostly graphene \cite{Zhang-2D}.

\begin{figure}[htpb]
\includegraphics[width=0.9\linewidth]{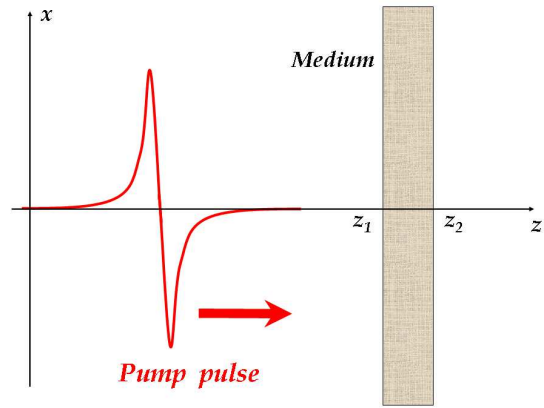}
\caption{(Color online) An optically thin layer of medium (shadowed area) is irradiated by a single-cycle pulse (red line) of plane wavefront propagating along the $z$ axis, which is orthogonal to the layer.
The pulse is linearly polarized along the $x$ axis.}
 \label{fig1}
\end{figure}

We start our consideratrion by assuming a plane layer of the medium is irradiated by a single-cycle pump pulse having plane wavefront and propagating along the $z$ axis, as shown
in Fig.~\ref{fig1}. The set of equations that we used in our study consists of the wave equation for the electric field $E(z,t)$ coupled with the equation for the response of the medium. 
We consider two types of the medium: dielectric, which is modeled as linear and elastically bounded electron oscillators, and metallic, treated as the free-running electrons in crystal lattice
within the framework of Drude model. The medium layer is assumed to be transversely uniform with the thickness $h = z_2-z_1$ and an incident single-cycle pulse is suggested linearly 
polarized along the $x$ axis and having the plane wavefront. This gives the following one-dimensional and scalar problem:

\begin{eqnarray}
\nonumber
\frac{\partial^2 E(z,t)}{\partial z^2}-\frac{1}{c^2}\frac{\partial^2 E(z,t)}{\partial t^2} = \frac{4\pi}{c^2}\Big( \frac{\partial^2 P(z,t)}{\partial t^2} + \\
\frac{\partial j(z,t)}{\partial t} \Big),
\label{eq_wave} \\
\ddot{x}_e+\gamma\dot{x}_e+\omega_{0}^2x_e=-\frac{e}{m_e}E(z,t),
\label{eq_disp} 
\end{eqnarray}
where $P(z,t)$ is the medium polarization related to the bounded charges, $j(z,t)$ is the current density of the free-charge carriers,  $x_e$ denotes the electron displacement  in the
medium, $\omega_{0}$ is the electron oscillations frequency, $\gamma$ is the damping rate and $m_e$ is the electron mass. The wave is considered to be linearly polarized along the
electron oscillation direction $x$ which is perpendicular to the propagation direction of a wave $z$. In the case of the dielectric medium the polarization is given as $P(z,t)=-N_0ex_e$,
where $N_0$ is the volume density of atoms in the medium, all the charge carriers are bounded in the atoms $j(z,t)=0$ and $\omega_{0}$ stands
for the medium resonance frequency. For the metallic medium we assume all the charge carriers are free and metal is electrically neutral, resulting in $P(z,t)=0$, $j(z,t)=-N_0e\dot{x}_e$
and $\omega_{0}=0$. In the latter case the well-known Drude model for the optical response of metals is restored from Eqs. (\ref{eq_wave})-(\ref{eq_disp}).

%%%%%%%%%%%%%%%%%%%%
%%%%%%%%%%%%%%%%%%%%

The corresponding Green's function of the one-dimensional wave operator is well-known and given as \cite{Tikhonov}:
\begin{eqnarray}
G(z,t)=-\frac{c}{2} \Theta(ct-|z|),
\label{Green_1D} 
\end{eqnarray}
where $\Theta$ is the Heaviside step function. For the dielectric medium we get only the first term on the right side of ~\eqref{eq_wave}. Considering firstly an extremely thin
layer of the medium $P(z,t)=\widetilde{P}(t) \delta(z)$, the field emitted by the dielectric medium is given as the convolution of the Green's function ~\eqref{Green_1D} with
the right hand of ~\eqref{eq_wave}:
\begin{eqnarray}
\nonumber
E_{\mathrm{out}}(z,t)= -\frac{c}{2} \int_{-\infty}^{t} \int_{-\infty}^{+\infty}  \frac{4\pi}{c^2} \frac{\partial^2 P(z',\tau)}{\partial \tau^2} \Theta \Big( c(t-\tau)-  \\
\nonumber
|z-z'| \Big) dz' d\tau = -\frac{2\pi}{c} \int_{-\infty}^{t}  \int_{z-c(t-\tau)}^{z+c(t-\tau)}  \frac{d^2 \widetilde P(\tau)}{d \tau^2} \cdot  \\
\nonumber
\delta(z') dz' d\tau = -\frac{2\pi}{c} \int_{-\infty}^{t} \frac{d^2 \widetilde P(\tau)}{d \tau^2} \Theta \Big( t - \tau - \frac{|z|}{c} \Big) d\tau = \\
\nonumber
-\frac{2\pi}{c} \dot{\widetilde P} \Big( t - \frac{|z|}{c} \Big). \\
\label{Field_1D} 
\end{eqnarray}

If the medium fits the layer placed between $z=z_1$ and $z=z_2$ we get from ~\eqref{Field_1D}:
\begin{equation}
E_{\mathrm{out}}(z,t)=-\frac{2\pi}{c} \int_{z_1}^{z_2} \frac{\partial P\Big( z',t-\frac{|z-z'|}{c}\Big )}{\partial t} dz'.
\label{Field_1D_thick} 
\end{equation}

Since $P(z,t) \sim x_e$, it is seen from ~\eqref{Field_1D_thick}, that the field emitted by the layer in this one-dimensional problem is determined by the velocity of the oscillating electrons in the medium.
It is important to note that the one-dimensional consideration is valid for the distances from the layer and the pulse wavelengths that are much less than the transverse size of the layer.
For a metal we obtain a similar expression:
\begin{equation}
E_{\mathrm{out}}(z,t)=-\frac{2\pi}{c} \int_{z_1}^{z_2}  j\Big( z',t-\frac{|z-z'|}{c}\Big ) dz'.
\label{Field_1D_thick-2} 
\end{equation}
\eqref{Field_1D_thick-2} implies that the field emitted by the metallic layer is proportional to the electron velocity in exactly the same way. It is in contrast to the well-known result
for the three-dimensional case when the field is measured at the distances from the layer much greater than its transverse sizes:
\begin{equation}
\nonumber
\overline{E}_{\mathrm{out}}(\overline{r}',t)\thicksim\ddot{\overline{P}}\Big( \overline{r},t-\frac{|\overline{r}-\overline{r}'|}{c} \Big ) \thicksim \dot{\overline{j}}\Big( \overline{r},t-\frac{|\overline{r}-\overline{r}'|}{c} \Big ).
\label{Field_3D} 
\end{equation}

The specific form of the layer response Eqs. (\ref{Field_1D})-(\ref{Field_1D_thick-2}) in the case of a linear medium offers the possibility to produce a unipolar half-cycle pulse under the excitation of the medium by a bipolar
pump pulse. Indeed, let us suppose that we have a pump pulse containing one optical cycle (that is, a single sign-changing oscillation of the field as in Fig.~\ref{fig1}) at the input. Then, under the influence of the first half of the pulse electrons in the medium will be accelerated from a standstill to a certain velocity.
When the driving electric field changes its sign at the second half of the pump pulse, the oscillating electrons will start decelerating. However, the velocity of electrons although decreasing will still keep the 
same sign, keeping also the sign of the emitted field ~\eqref{Field_1D}. As a result, the total emission of the medium will have the form of a half-cycle pulse.

To illustrate the above reasoning we performed numerical simulations of the system Eqs. (\ref{eq_wave})-(\ref{eq_disp}).
We consider the excitation pulse having a symmetric shape of the following form:
\begin{equation}
E_{\mathrm{in}}(z,t)=E_0e^{-(t-\frac{z}{c} )^2/\tau_{p}^2}\sin \Omega \Big(t-\frac{z}{c} \Big),
\label{eq9} 
\end{equation}
where $\Omega$  is the central frequency and $\tau_p$ is the pulse duration. We assume the incident pulse $E_{\mathrm{in}}(z,t)$ to have a single-cycle shape. That is, the pulse 
contains only one cycle of optical oscillations within the pulse duration $\tau_p$.

\begin{figure}[htpb]
\includegraphics[width=0.9\linewidth]{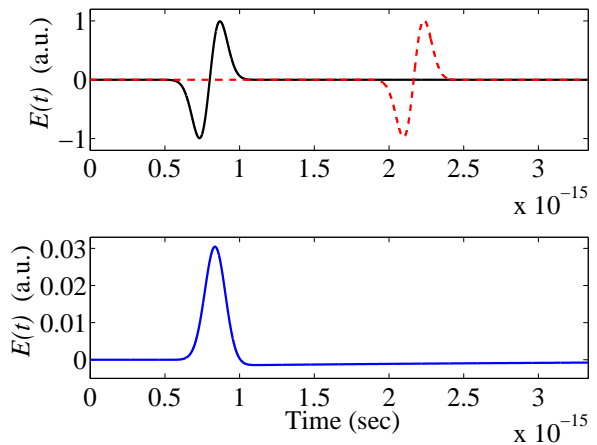}
\caption{(Color online) Transmitted pulse (top panel, red dashed line) and reflected pulse (bottom panel), obtained after a single-cycle bipolar pulse ~\eqref{eq9} (top panel, black solid line) passing through 
an optically thin metallic layer; the layer thickness $h = 1$ nm, other parameters are $N_0 = 5 \cdot 10^{22}$ cm$^{-3}$, $\gamma = 10^{13}$ sec$^{-1}$, $\Omega = 6.3 \cdot 10^{15}$ sec$^{-1}$,
$\Omega \tau_p = 0.1$.}
\label{fig2}
\end{figure}

Fig.~\ref{fig2} shows the results of the simulations for a single-cycle bipolar pulse passing through an optically thin metallic layer where the amplitude of the incident pulse
is scaled to unity. 

%%%%%%%%%%%%%%%%%%%%
%%%%%%%%%%%%%%%%%%%%

\begin{figure}[htpb]
\includegraphics[width=0.9\linewidth]{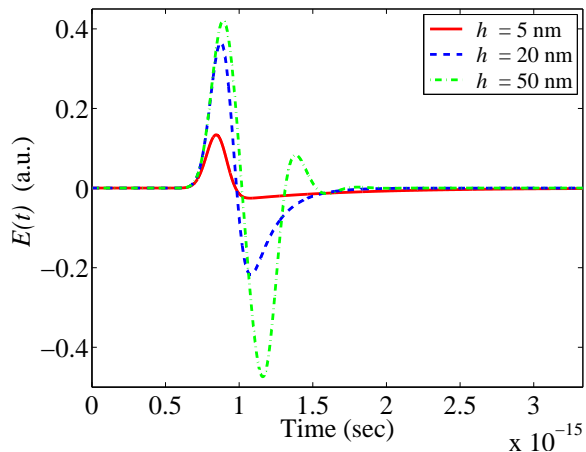}
\caption{(Color online) Pulse obtained after reflection of a single-cycle pulse with the parameters as in Fig.~\ref{fig2} from a thin metallic layer in dependence on its thickness; other parameters are the same as in Fig.~\ref{fig2}.}
 \label{fig3}
\end{figure}

It is seen from Fig.~\ref{fig2} that the reflected pulse has a leading peak of the constant sign accompanied by a weak and long tail after the trailing edge. The maximum amplitude of the reflected 
pulse is by about two orders of magnitude smaller than that of the incident pulse due to the small thickness of the layer. Thickness of the layer is of crucial importance for the whole process 
since the unipolar field emitted by the outer parts of the medium transforms the incident pulse adding the zero-frequency spectral component. Then, the inner parts are excited by the pulse
of modified spectral content that is not already bipolar. As a result, the reflected pulse contains a long tail of different sign. Fig.~\ref{fig3} illustrates the reflected pulse transformation
 when increasing the thickness of the layer. It is seen that the tail rapidly grows with the layer thickness and becomes comparable in amplitude with the leading peak. That is, both reflected and
 transmitted pulses become essentially bipolar as the thickness of the layer increases.

%%%%%%%%%%%%%%%%%%%%
%%%%%%%%%%%%%%%%%%%%

\begin{figure}[htpb]
\includegraphics[width=0.9\linewidth]{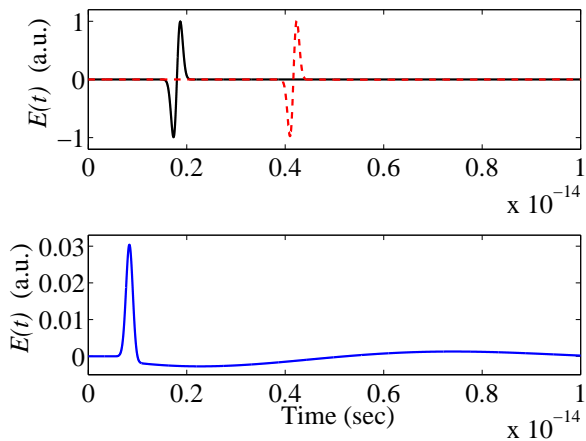}
\caption{(Color online) Transmitted (top panel, red dashed line) and reflected pulses (bottom pane), obtained after reflection of a single-cycle bipolar pulse ~\eqref{eq9} (top panel, black solid line)
from a thin dielectric layer; $ \Omega \tau_p = 0.1, \omega_{0}=\Omega/10$, other parameters are the same as in Fig.~\ref{fig2},~\ref{fig3}.}
\label{fig4}
\end{figure}

For the case of a dielectric medium the situation is rather similar as it is illustrated in Fig.~\ref{fig4}. Reflected pulse has now the form of a short unipolar peak accompanied by a long weak tail.
In contrast to the previous case the tail is oscillating and results from the decay of the dipoles. 

%%%%%%%%%%%%%%%%%%%%
%%%%%%%%%%%%%%%%%%%%

It is worth paying attention to the value of the constant spectral component of the reflected pulses. 
Indeed, according to ~\eqref{Field_1D_thick} in the case of the dielectric medium:
\begin{eqnarray}
\nonumber
\int_{-\infty}^{+\infty} E_{\mathrm{out}}(z,t')dt' \sim \\
\nonumber
\int_{z_1}^{z_2} dz' \int_{-\infty}^{+\infty} \frac{\partial P\Big( z',t'-\frac{|z-z'|}{c}\Big )}{\partial t'} dt' = \\
\nonumber
 \int_{z_1}^{z_2} \Big[ P(z',+\infty) - P(z',-\infty) \Big] dz' = 0, \\
\label{Eq_area_diel} 
\end{eqnarray}
since the electrons are assumed to be at the standstill long before and long after the action of the pump pulse. As a result, the total time integral of the
electric field over the whole pulse duration vanishes, as required for a propagating pulse. However, the most applications do not 
require the pulses to have the strictly constant sign of the electric field all along. It is enough that the pulses are "approximately unipolar". It is interesting to note, however, that if the intensive excitation pulse can cause the
irreversible changes of the medium optical properties, so that it leaves the medium having permanent dipole moments, ~\eqref{Eq_area_diel} will not be satisfied and the emitted 
pulse can thus be strictly unipolar. It is important to stress that there is no contradiction with the principle of the pulse area conservation \cite{Rosanov-2} since this principle
does not impose any restrictions on the possibility of the pulse area gain by the excited medium.

For the metallic layer ~\eqref{Eq_area_diel} can be broken, since the electrons are considered free and thus get the certain nonzero displacement
under the action of a single-cycle pulse:
\begin{eqnarray}
\nonumber
\int_{-\infty}^{+\infty} E_{\mathrm{out}}(z,t')dt' \sim \int_{z_1}^{z_2} dz' \int_{-\infty}^{+\infty} j\Big( z',t'-\frac{|z-z'|}{c}\Big ) dt' \\
\nonumber
= Q_\Sigma,
\label{Eq_area_metal} 
\end{eqnarray}
where $Q_\Sigma$ denotes the total charge having passed through the transverse section of the layer orthogonal to the $x$ axis. This result, however, seems to represent the one-dimensional nature of the problem.
 In a metallic layer with finite transverse sizes such displacement of free electrons will lead to the formation of uncompensated charge at the
layer boundaries and thus the medium will eventually relax to the electrically neutral state with $Q_\Sigma=0$. This would lead again to a quasi-unipolar pulse with a long small-amplitude tail fully compensating the area of the leading peak.

To summarize, we have proposed theoretically a method for unipolar half-cycle pulse generation using unusual properties of simple linear reflection from a thin layer of a metallic or dielectric
medium irradiated by a single-cycle pulse. The specific feature of the considered one-dimensional geometry is that emitted field appears to be determined by the velocity of oscillating charges (and not by their acceleration)
due to the form of the Green's function of the one-dimensional wave operator. This result seems to conflict with common views and radically differs from the usual mirror reflection of long pulses what opens up further new possibilities for controlling few-cycle pulses.  

In the case of metallic layer and a single-cycle incident pulse the velocity of free charges can be forced to keep its sign resulting in the unipolar half-cycle reflected pulse. However, emitted pulses were found to be strictly unipolar provided that the displacement of charges does not relax back to zero after the pump pulse passage and with neglect of the finite layer thickness and the oscillators decay. In fact, the pulses we obtain are approximately unipolar, that is, the main part of the pulse having one and the same sign. In particular, making allowance for the pulse propagation through a metallic layer of finite thickness and the decay of the free charge carriers, the emitted pulse will possess the weak tail of opposite sign behind the trailing edge of the pulse which can be obtained, however, to be arbitrarily small compared to the leading peak. In the case of the charges being bounded in the dielectric medium, the emitted field will contain the long and weak oscillating tail so that the integral of the field over the whole pulse vanishes. 

Our approach for the unipolar half-cycle pulse reflection is applicable for the central wavelengths of the incident pulse that are much less than the lateral dimensions of the medium layer. That is why the proposed method can be principally applied in the wide frequency range of the excitation pulses. We would like to especially emphasize that our method can be potentially extended to the terahertz region where it is easier to generate the single-cycle pulses \cite{Fattinger, Bartel, Hoffmann}.

\section*{Funding Information}

\textbf{Funding.} Russian Foundation for Basic Research (16-02-00762) and Government of Russian Federation, Grant 074-U01; I.B. thanks German Research Foundation (DFG) (project BA 4156/4-1); Nieders. Vorab (project ZN3061).

%\bibstyle{apsrev4-1.bst}
\bibliography{optics}

\end{document}